# A Cascade Electron Source Based on Series Horizontal Tunneling Junctions

Zhiwei Li and Xianlong Wei

*Abstract*—On-chip electron sources have wide potential applications in miniature vacuum electronic devices and emission efficiency is one of their performance benchmarks. A cascade electron source based on series metal-insulator-metal horizontal tunneling junctions is proposed, where free electrons are additively extracted from each tunneling junction. A cascade electron source with $n$ horizontal tunneling junctions shows a theoretical emission efficiency of approximately $\eta(n) = 1 - (1 - \eta_0)^n$, with $\eta_0$ being the efficiency of a single tunneling junction. Experimentally, a cascade electron source with three Si-SiO$_x$-Si tunneling junctions is demonstrated, achieving an emission efficiency as high as 47.6%. This work provides a new way of realizing highly efficient on-chip electron sources.

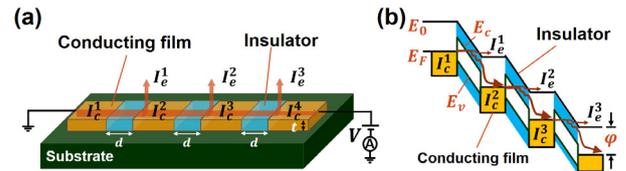

**Fig. 1** Structure and working principle of a CES. **(a)** Schematic diagram of a CES with three identical HTJs in series. **(b)** The energy band diagram of the CES in **(a)**. $E_F$, $E_0$, $E_v$ and $E_c$ are Fermi level, vacuum level, valence band top and conduction band bottom, respectively.

## I. Introduction

ELECTRON sources fabricated on a chip have attracted continuous interest over a long time due to their promising applications in miniature vacuum electronic devices and systems [1-6]. Compared with widely studied field emission sources based on tip arrays [6], tunneling electron sources based on metal-insulator-metal (MIM) or metal-oxide-semiconductor (MOS) structures show higher potential for on-chip applications for the benefits of lower operating voltage, lower operating vacuum and easier fabrication [1, 4]. Electron emission from these electron sources is attributed to the escape of hot electrons generated in electrically-biased MIM/MOS structures. Emission efficiency, which is defined as the ratio of electron emission current to the conduction current of MIM/MOS structures, is therefore a benchmark for the electron source performances. However, the efficiency is typically less than 1% for several decades because of the drastic energy dissipation of hot electrons across the top metal layer. These years, the emission efficiency has been successfully boosted to over 10% through several different ways [1, 4, 7]. By using graphene as the top layer of a MOS structure to suppress the hot electron dissipation, an emission efficiency of up to 32.1% has been reported [4].

Recently, we proposed a high-efficiency electron source based on a horizontal tunneling junction (HTJ), where an insulator nanogap is horizontally sandwiched between two thin conducting films on an insulating substrate [8]. It shows a theoretical emission efficiency of up to 21.0-25.9%, and an experimental value of 16.6% by using a Si-SiO$_x$-Si HTJ in electroformed silicon oxide [7]. In contrast with conventional tunneling electron sources with vertically-stacked MIM or MOS structure where electrons are emitted from the surface of the topmost layer [1, 4], the unique architecture of HTJ enables lateral electron extraction from MIM or MOS structure. This makes it possible to additively extract free electrons from series HTJ in a cascade way.

Here, we propose a cascade tunneling electron source consisting of series HTJs and analyze its emission efficiency in theory. Based on three Si-SiO$_x$-Si HTJs in series, a cascade tunneling electron source is experimentally realized and an emission efficiency of up to 47.6% is demonstrated.

## II. Device Structure and Working Principles

The schematic structure of a cascade electron source (CES) is shown in Fig. 1a, where several (three in the diagram) MIM HTJs are connected in series on an insulating substrate. Each HTJ consists of a thin insulator nanogap horizontally connected between two thin conducting films. The thickness $t$ of the insulator nanogap and conducting film is set to less than mean free path of inelastic scattering. When a bias voltage larger than $\varphi/e$ ($\varphi$ is the work function of conducting film and $e$ is elementary charge) is applied to such an HTJ, electrons tunnel from low-potential conducting film to the insulator nanogap. Some of them are then accelerated to above the vacuum level $E_0$ by the electric field in the insulator (Fig. 1b) and escape laterally from MIM junction [7]. The predicted efficiency of such an HTJ reaches as high as 25.9% [8].

When a bias voltage is applied on the two outmost electrodes of this CES, electron emission from all HTJs can take place simultaneously and emission current from each HTJ is additive due to the lateral electron extraction (Fig. 1b).

This work was supported in part by the National Key Research and Development Program of China (Grant No. 2019YFA0210201, 2017YFA0205003), NSF of China (Grant No. 11874068).

Zhiwei Li, and Xianlong Wei are with the key Laboratory for the Physics and Chemistry of Nanodevices, Department of Electronics, Peking University, Beijing 100871, P.R. China (E-mail: weixl@pku.edu.cn).

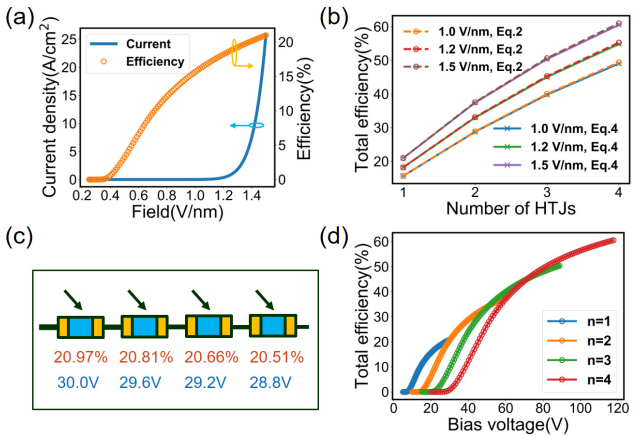

**Fig. 2** Theoretical simulation of CESs based on Si-SiO$_x$-Si HTJs. **(a)** Conduction current and emission efficiency of a single Si-SiO$_x$-Si HTJ versus the electric field strength. **(b)** Emission efficiency of CESs versus the number of HTJs at different $E^1$. Eq. (2) and Eq. (4) are plotted. **(c)** The voltage-drop and emission efficiency of each Si-SiO$_x$-Si HTJ for a CES with $n$=4. Both the voltage-drop and efficiency drop slightly step by step from the first HTJ (left) to the last one (right). **(d)** Emission efficiency of CESs versus bias voltage.

Assuming the same emission efficiency of $\eta_0$ for all HTJs, we get the conduction current and emission current of the $i$th HTJ as $I_c^i = I_c^1(1-\eta_0)^{i-1}$ and $I_e^i = I_c^i \eta_0 = I_c^1(1-\eta_0)^{i-1}\eta_0$, respectively, where $I_c^1$ is conduction current entering the first HTJ. The total emission current of a CES can be obtained as
$$I_e = \sum_{i=1}^{n} I_c^1(1-\eta_0)^{i-1}\eta_0 = I_c^1 - I_c^1(1-\eta_0)^n \quad (1)$$
. The total emission efficiency of a CES is then obtained as
$$\eta = \frac{I_e}{I_c^1} = 1 - (1-\eta_0)^n \quad (2)$$
Therefore, emission efficiency $\eta$ of a CES increases with the number $n$ of series HTJs. In case of $\eta_0 \ll 1$, we can obtain $\eta \approx n\eta_0$, which means $\eta$ increases approximately linearly with $n$. As shown in Fig. 2b, $\eta$ gets a value of 61.0% when $n = 4$ and $\eta_0 = 21.0\%$, demonstrating good effectiveness of CESs in improving emission efficiency.

### III. THEORETICAL SIMULATION

To further illustrate the performance of a CES, its emission efficiency is theoretically simulated with specific Si-SiO$_x$-Si HTJs. Si conducting filament was found to form in the shallow surface of electroformed silicon oxide substrate through a SiO$_x$ → Si electrochemical reduction process, and can be reversibly ruptured and connected under voltage stress [9]. The rupture of a conducting filament results in a Si-SiO$_x$-Si HTJ in shallow surface of the substrate, which was thought to be responsible for electron emission from silicon oxide electron-emitting diodes [7]. Conduction current of a Si-SiO$_x$-Si HTJ at a bias voltage of $V$ can be described by Fowler-Nordheim law [10]
$$I(E) = A\frac{e^3 E^2}{8\pi h\chi}\exp\left(-\frac{8\pi(2m^*)^{1/2}\chi^{3/2}}{3heE}\right) \quad (3)$$
, where $A$ is the cross section area of a HTJ, $h$ is the Planck's constant, $\chi$ is the potential barrier at the Si-SiO$_x$ interface, $m^*$ is the effective mass of SiO$_x$, and $E = V/d$ is the strength of electric field, with $d$ being the width of a HTJ. According to the method in our previous work [8], the curve of $\eta(E)$ together with that of $I(E)$ is calculated as shown in Fig. 2a for a Si-SiO$_x$- Si HTJ with the parameter values of $A$=4 nm×10 nm, $\chi$=3 eV, $m^*$ =0.42$m_0$, and $d$=20 nm, where $m_0$ is the rest mass of an electron.

Emission efficiency of a CES with $n$ identical Si-SiO$_x$-Si HTJs can then be simulated based on the relations of $\eta(E)$ and $I(E)$ in Fig. 2a. Despite the same geometric structure, each HTJ in the CES will show a different conduction current due to the extraction of emission current from the circuit, and thus different voltage-drop and emission efficiency for each HTJ. If the voltage-drops across conducting filaments are neglected, self-consistent equations for a CES with $n$ identical Si-SiO$_x$-Si HTJs can be obtained:
$$\begin{cases} V = \sum_{i=1}^{n} V^i \\ E^i = \frac{V^i}{d}, i = 1,2,\dots,n \\ I_c^i = I(E^i), i = 1,2,\dots,n \\ I_c^{i+1} = I_c^i\left(1-\eta(E^i)\right), i = 1,2,\dots,n \\ I_e = I_c^1 - I_c^{n+1} \\ \eta = \frac{I_e}{I_c^1} \end{cases} \quad (4)$$
, where $V$ is the voltage applied to the CES, $V^i$ and $E^i$ is the voltage-drop and the electric field strength of the $i$th HTJ, $I_c^{n+1}$ is conduction current going out of CES, $\eta(E)$ and $I(E)$ are the functions plotted in Fig. 2a. The 1st and 5th equations are derived from Kirchhoff's laws, and the 4th and 6th equations are derived directly from the definition of emission efficiency. There are $3n + 3$ equations and $3n + 3$ unknown variables ($V^i, E^i, I_c^i, I_c^{n+1}, I_e$, and $\eta$), so the equations can be numerically solved with unique solutions for a given $V$ and $d$.

Fig. 2b shows the calculated $\eta$ of a CES with respect to $n$ at different $E^1$ by Eq. (4) (solid lines). It can be seen that $\eta$ increases with $n$ for a fixed $E^1$. When $E^1$=1.5 V/nm, $\eta$ has the value of 21.0%, 37.4%, 50.3%, 60.5%, respectively, for n=1 to 4. For a fixed $n$, $\eta$ also increases with $E^1$. A CES with four HTJs has emission efficiency of 49.0%, 54.8%, and 60.5%, respectively, for $E^1$=1.0, 1.2 and 1.5 V/nm. We also plot $\eta$ with respect to $n$ in Fig. 2b by taking $\eta_0 = \eta(E^1)$ into Eq. (2) (dotted lines). Self-consistently calculated efficiencies from Eq. (4) are slightly lower (<0.5%) than that from Eq. (2). Therefore, Eq. (2) can give $\eta$ of a CES with identical HTJs in a good approximation. The reason for the high consistency is attributed to the exponential dependence of conduction current on the strength of electric field in Fowler-Nordheim equation (Fig. 2a), which means a considerable variation of $I_c^i$ resulted from electron emission will not bring much variation in the strength of electric field $E^i$ or emission efficiency $\eta(E^i)$ of neighboring HTJs. Fig. 2c shows voltage-drops ($V^i$) and emission efficiencies ($\eta(E^i)$) of each HTJ in a CES with $n$=4. The quite similar emission efficiency for the four HTJs further confirms that it is reasonable to assume the same emission efficiency as in obtaining Eq. (2). Fig. 2d shows the dependence of $\eta$ on $V$. $\eta$ increases with $V$ until reaching a maximum value when $E^1$ increases to 1.5 V/nm, the maximum electric field strength that silicon oxide can withstand [11]. It can be seen that the maximum achievable $\eta$ of a CES increases with $n$.

## IV. Experimental Demonstration

A CES with three Si-SiO$_x$-Si HTJs is experimentally demonstrated. Fig. 3a shows a schematic diagram of a CES with three Si-SiO$_x$-Si HTJs and its measurement setup. A scanning electron microscopy (SEM) image of a CES before measurement is shown in Fig. 3b, where three Au/Ti nanowires (Au/Ti =15/5 nm, 2 μm in length and 200 nm in width) are fabricated in series on a SiO$_2$ substrate (SiO$_2$/Si=300 nm/525 μm) with each Au/Ti nanowire connected between two electrodes (Au/Ti =45/5 nm). Emission current measurement is carried out in a probe station (JANIS) with a vacuum level of $10^{-2}$~$10^{-3}$ Pa. A tungsten probe around 200 μm above the device with a bias of 210 V is used to collect emission current. The electron emission performances and electrical transport properties are measured with a KEITHLEY 4200 semiconductor system in direct current mode.

To form a Si-SiO$_x$-Si HTJ, a Ti/Au nanowire is firstly broken by Joule heating under a low voltage of around 1.5 V to form a nanogap (Fig. 3b, inset), and then SiO$_2$ under the broken gap is electroformed [9]. Electroformed SiO$_2$ exhibits a unipolar resistive switching behavior (Fig. 3c), which is attributed to the reversible rupture and connection of Si conducting filament formed in shallow surface of substrate [7, 9]. A Si-SiO$_x$-Si HTJ is therefore formed in the nanogap when a conducting filament is ruptured in high-resistance state (Fig. 3a). A Si-SiO$_x$-Si HTJ can generate considerable electron emission [7], as shown in Fig. 3c. Emission current (orange curve) of a Si-SiO$_x$-Si HTJ rises with the increasing bias voltage until reaching 2.57 μA, and the maximum emission efficiency reaches 15.9% (2.57 μA/16.2 μA), which is comparable with that reported in Ref. [7].

After a Si-SiO$_x$-Si HTJ is formed between each pair of neighboring electrodes, a CES with three Si-SiO$_x$-Si HTJs is prepared. Its electron emission is driven by applying a bias voltage to the two outmost electrodes. Since the generation and rupture of conducting filaments are drastic redox processes, the local structure of HTJs in the CES changes constantly, bringing large fluctuations to the curves of conduction current and emission current (Fig. 3d). The conduction current undergoes a fast reduction at high-voltage region for the same reason. Comparatively, the emission current increases faster than the conduction current, giving rise to an increasing emission efficiency. An emission current of 5.0 μA is achieved at 57.1 V for a CES with three Si-SiO$_x$-Si HTJs, corresponding to an emission efficiency as high as 47.6% (5.0 μA/10.5 μA) (Fig. 3d). The efficiency is much higher than that (32.1%) of previous graphene-oxide-semiconductor tunneling electrons, which uses atomic graphene as the top electrode to minimize the dissipation of hot electrons [4]. It agrees well with the theoretical estimate of 50.3% ($\eta_0$=21.0%, $E^1$=1.5 V/nm, $n$=3) for a CES with three HTJs in last section.

An SEM image of the CES after tens of emission current measurements is shown in Fig. 3e, where some contaminations can be clearly seen between each pair of neighboring electrodes. Electron-beam-induced deposition of amorphous carbon is a commonly seen phenomenon in instruments with free electron beams when it does not operate in high vacuum [12]. Since the device before measurement is clean as shown in Fig. 3b, the contamination is thought to be amorphous carbon deposition arising from the radiolysis of residual hydrocarbon molecules by free electrons emitted from HTJs. The presence of amorphous carbon between each pair of neighboring electrodes provides a solid evidence that electrons are emitted from the three Si-SiO$_x$-Si HTJs in the CES at the same time. Though the three HTJs may not show identical structures as in the theoretical model, which is speculated from quite different amount of amorphous contamination around them, our measured efficiency as high as 47.6% unambiguously indicates that CESs provide an effective way of boosting emission efficiency.

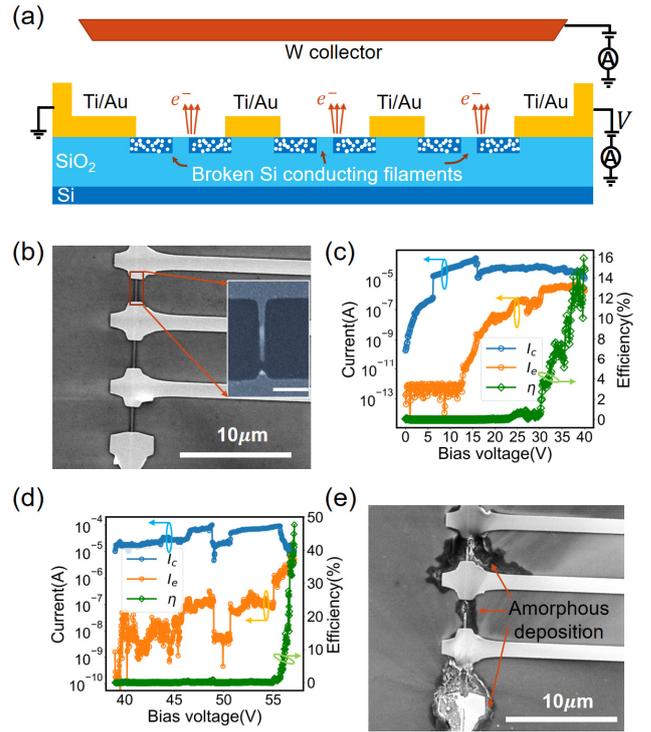

**Fig. 3** Experimental demonstration of a CES based on Si-SiO$_x$-Si HTJs. **(a)** Schematic structure of a CES with 3 Si-SiO$_x$-Si HTJs and its measurement setup. Each Si-SiO$_x$-Si HTJ is formed by a broken Si conducting filament. **(b)** SEM image of a CES with 3 Si-SiO$_x$-Si HTJs. The inset shows the SEM image of a broken gold nanowire. **(c-d)** Simultaneously measured conduction current (blue curve) and emission current (orange curve) of a Si-SiO$_x$-Si HTJ (c) and a CES with 3 Si-SiO$_x$-Si HTJs (d). The efficiency is also plotted as a green curve in (c) and (d). **(e)** SEM image of the CES in (c) after electrical measurement. The device is destroyed by high electric field in the end, leaving lots of amorphous carbon deposition around the emitting area.

## V. Conclusion

A cascade electron source based on series HTJs is proposed for the first time as a new form of on-chip electron sources. Through both theoretical simulation and experimental demonstration, CES is proved to be an effective way to further improve the emission efficiency of tunneling electron sources. Based on Si-SiO$_x$-Si HTJs formed in electroformed SiO$_2$, a cascade electron source with three HTJs is demonstrated in experiments, and an emission efficiency as high as 47.6% has been achieved. Our work provides a new way of realizing highly efficient on-chip electron sources.